\documentclass[journal,comsoc,10pt]{IEEEtran}

\usepackage[T1]{fontenc}

\usepackage{cite}
\usepackage[colorlinks,
linkcolor=red,
anchorcolor=blue,
citecolor=green
]{hyperref}

\ifCLASSINFOpdf
\else
\fi
\usepackage{amsmath}
\usepackage{diagbox}

\usepackage{ amssymb }
\usepackage{amsfonts}
\usepackage{caption}
\usepackage{graphicx}
\usepackage{lettrine}
\usepackage{epstopdf}
\usepackage{textcomp,booktabs}
\usepackage[usenames,dvipsnames]{color}
\usepackage{colortbl}
\definecolor{mygray}{gray}{.9}
\definecolor{mypink}{rgb}{.99,.91,.95}
\definecolor{mycyan}{cmyk}{.3,0,0,0}
\usepackage{pifont}
\usepackage{subfigure}
\usepackage{multirow}
\usepackage{url}
\usepackage{color}
\usepackage{threeparttable}
\usepackage{setspace}
\usepackage{lineno}
\usepackage{tabularx}
\usepackage[dvipsnames,usenames]{color}

\usepackage{bm}

\usepackage{algorithm} 
\usepackage{algorithmic} 

\usepackage{dblfloatfix}

\usepackage[dvipsnames,usenames]{color}
\usepackage[usenames,dvipsnames]{color}

\definecolor{light-gray}{gray}{0.90}
\begin{document}

	\title{Semantic Satellite Communications Based on Generative Foundation Model}

	\author{Peiwen Jiang,~\IEEEmembership{Graduate~Student~Member,~IEEE,} Chao-Kai Wen,~\IEEEmembership{Fellow,~IEEE,}\\ Xiao Li,~\IEEEmembership{Member,~IEEE,} Shi Jin,~\IEEEmembership{Fellow,~IEEE,} and Geoffrey Ye Li,~\IEEEmembership{Fellow,~IEEE}
			\thanks{P. Jiang, X. Li and S. Jin are with the National
				Mobile Communications Research Laboratory, Southeast University, Nanjing
				210096, China (e-mail: PeiwenJiang@seu.edu.cn; li\_xiao@seu.edu.cn; jinshi@seu.edu.cn).}
			\thanks{C.-K. Wen is with the Institute of Communications Engineering, National
				Sun Yat-sen University, Kaohsiung 80424, Taiwan (e-mail: chaokai.wen@mail.nsysu.edu.tw).}
			\thanks{G. Y. Li is with the Department of Electrical and Electronic Engineering,
				Imperial College London, London SW7 2AZ, UK (e-mail: geoffrey.li@imperial.ac.uk).}}
	
	\maketitle
	\pagestyle{empty}  
	\thispagestyle{empty} 
%
%

\begin{abstract}

Satellite communications can provide massive connections and seamless coverage, but they also face several challenges, such as rain attenuation, long propagation delays, and co-channel interference. To improve transmission efficiency and address severe scenarios, semantic communication has become a popular choice, particularly when equipped with foundation models (FMs). In this study, we introduce an FM-based semantic satellite communication framework, termed FMSAT. This framework leverages FM-based segmentation and reconstruction to significantly reduce bandwidth requirements and accurately recover semantic features under high noise and interference. Considering the high speed of satellites, an adaptive encoder-decoder is proposed to protect important features and avoid frequent retransmissions. Meanwhile, a well-received image can provide a reference for repairing damaged images under sudden attenuation. Since acknowledgment feedback is subject to long propagation delays when retransmission is unavoidable, a novel error detection method is proposed to roughly detect semantic errors at the regenerative satellite. With the proposed detectors at both the satellite and the gateway, the quality of the received images can be ensured. The simulation results demonstrate that the proposed method can significantly reduce bandwidth requirements, adapt to complex satellite scenarios, and protect semantic information with an acceptable transmission delay.

\end{abstract}



\section{Introduction}
\IEEEPARstart{S}{atellite} communications (SatComs) have emerged as a promising solution for achieving massive connections and seamless coverage, essential for next-generation wireless networks. Unlike terrestrial communications, SatComs involve high launch costs, depend on line-of-sight (LoS) for information transmission, and entail long propagation distances. To mitigate transmission delays and provide high throughput at lower costs, several low Earth orbit (LEO) satellite systems, such as Iridium, Globalstar, OneWeb, Starlink, and Telesat \cite{di2019ultra}, have been deployed. Initially featuring the transparent type that functions as a signal amplifier, the regenerative type now plays a pivotal role in satellite-terrestrial integrated communications \cite{kodheli2020satellite,chen2020system}. A regenerative satellite, acting as a base station (BS) of the terrestrial network through on-board processing like modulation and coding, also supports inter-satellite links in the LEO satellite constellation, a key technique for global networking.

Addressing performance degradation due to the high mobility of LEOs, numerous studies have focused on enhancing the transmission reliability and adaptability of SatComs \cite{heo2023mimo}, including methods such as robust beamforming \cite{lu2020robust}, adaptive modulation \cite{agarwal2019analysis}, and interference mitigation \cite{nguyen2022outage}. The large constellation of LEO satellites and the diversity of services from wide coverage areas contribute to the complexity of transmission scenarios. Additionally, an aggressive frequency reuse scheme, often adopted in satellite systems for improved spectrum efficiency over a multiple color frequency reuse pattern, makes co-channel interference (CCI) a significant challenge \cite{vazquez2016precoding}. Efforts to address the adverse effects of CCI have led to extensive research on interference detection and mitigation \cite{peng2022integrating,cottatellucci2006interference,ding2017spectrum}. Nonetheless, the rapid movement of satellites renders the interference somewhat unpredictable. To ensure successful transmission, techniques such as the hybrid automatic repeat request (HARQ) are sometimes employed but are hindered by substantial and varied transmission delays (typically 2--20 ms between satellite and ground) \cite{ahmad2015enhancing,hosseinian2021review,qi2021standardization}. Consequently, deep learning has been proposed as a solution to these complex issues in SatComs \cite{fourati2021artificial}.

Semantic communication has demonstrated its potential in bandwidth reduction and resilience to poor channel conditions through a shared knowledge base (KB), effective semantic extraction, and reconstruction capabilities \cite{gunduz2022beyond,lu2021reinforcement,jiang2022deep,10158526}. Recent advancements have applied semantic communication to address specific challenges in the non-terrestrial network (NTN), including satellite computation offloading \cite{10445211}, wireless UAV control \cite{xu2023task}, and remote spectrum sensing \cite{yi2022semantic}. However, further exploration is needed to enhance satellite transmission performance through semantic communications. Presently, generative foundation models (FMs), such as large language models (LLMs), have shown remarkable capabilities in understanding and generating human-like language, as well as handling multimodal sources like images and audios. The integration of FMs into semantic communication \cite{jiang2023large} can significantly boost its performance and efficiency. For instance, GPT has been used to orchestrate multiple AI models within a communication system \cite{shen2023large}, and a semantic segmentation method has been developed to adaptively segment content using text control in \cite{wang2023seggpt}. Additionally, a diffusion model has been utilized to restore transmitted semantic features, even with only outlines of required objects, in \cite{grassucci2023generative}. While FM-based semantic communication presents an innovative approach, its application within SatComs necessitates further exploration.

In response to SatCom challenges, this study introduces a novel FM-based semantic satellite communication system (FMSAT). In FMSAT, we utilize SegGPT \cite{wang2023seggpt}  and the conditional diffusion model \cite{dhariwal2021diffusion} for image transmission from a user terminal (UT) to the gateway to reduce the required bandwidth. The system consists of a semantic encoder-decoder and an OFDM transceiver similar to the settings of Starlink \cite{humphreys2023signal}. SegGPT extracts features that the semantic encoder codes. The regenerative SatComs transmit the codeword through the OFDM transceiver, and the diffusion model reconstructs the received features. The proposed framework employs SegGPT and the diffusion model for dynamic extraction and reconstruction to cope with the rapidly changing channel conditions of SatComs. The formidable capabilities of FMs enable the extraction and safeguarding of essential parts of the transmitted image, alongside the rectification of transmission errors under challenging SNR conditions and substantial interference.

Furthermore, to counteract sudden deep attenuations due to rapid satellite movement, FM-based reconstruction exploits content correlation, enabling the repair of fragmented images through references to previously transmitted images. Nonetheless, when key features are obliterated by unpredictable interference, the FM-based reconstruction method also falters. In such instances, retransmission represents a feasible solution, though it may lead to substantial transmission delays. To tackle this issue, we suggest incorporating a signal processing component within the regenerative satellite, contingent upon available computational resources. Through effective feedback mechanisms, our framework incorporates two-tiered error detectors: an initial, rough detector assesses key feature errors without involving the complex semantic encoder-decoder at the satellite level, followed by a more detailed semantic loss assessment at the gateway to ensure transmission integrity.

Our contributions are summarized as follows:
\begin{itemize}
\item \textbf{Novel semantic framework for SatComs:} The proposed FMSAT adapts to the complex channel conditions induced by satellite movement and frequency reuse, leveraging FMs' flexibility to adjust transmitted features and reconstruct damaged sections, accommodating rapid SNR variations and CCI. The adaptability extends from transmitting selective semantic features to the complete image, ensuring consistently high-quality image reception.

\item \textbf{Exploiting channel condition and content correlation:} Different elements influencing SatComs' transmission errors are considered, allowing for an adaptive semantic encoder-decoder architecture. Compact additional encoders at the UT protect essential features, while various decoders at the gateway restore specific features. The extraction of similar background knowledge from previously transmitted images by FMs further enhances reconstruction under suboptimal channel conditions.

\item \textbf{Semantic error detectors for regenerative satellite:} Given the computational limitations of regenerative satellites and the associated transmission delays, a dual-level error detection system is proposed. The satellite's initial rough error detection rapidly identifies feature errors, triggering ACK/NACK feedback to the UT. Subsequently, the gateway's detailed error detection ensures the quality of transmitted content.
\end{itemize}

The rest of this paper is organized as follows. Section \uppercase\expandafter{\romannumeral2} introduces the system model, including the conventional SatCom architecture and the channel model. The proposed transmission frameworks are presented in Section \uppercase\expandafter{\romannumeral3}, and their detailed architectures and training processes are discussed in Section \uppercase\expandafter{\romannumeral4}. Section \uppercase\expandafter{\romannumeral5} demonstrates the superiority of the proposed networks under varying channel conditions and changing user requirements, along with their ability to save resources and reconstruct poor semantic features. Finally, Section VI concludes the paper.

\section{System Model}

\subsection{Regenerative SatComs}

\begin{figure}[h]
	\centering

  {

  \includegraphics[width=0.99\linewidth]{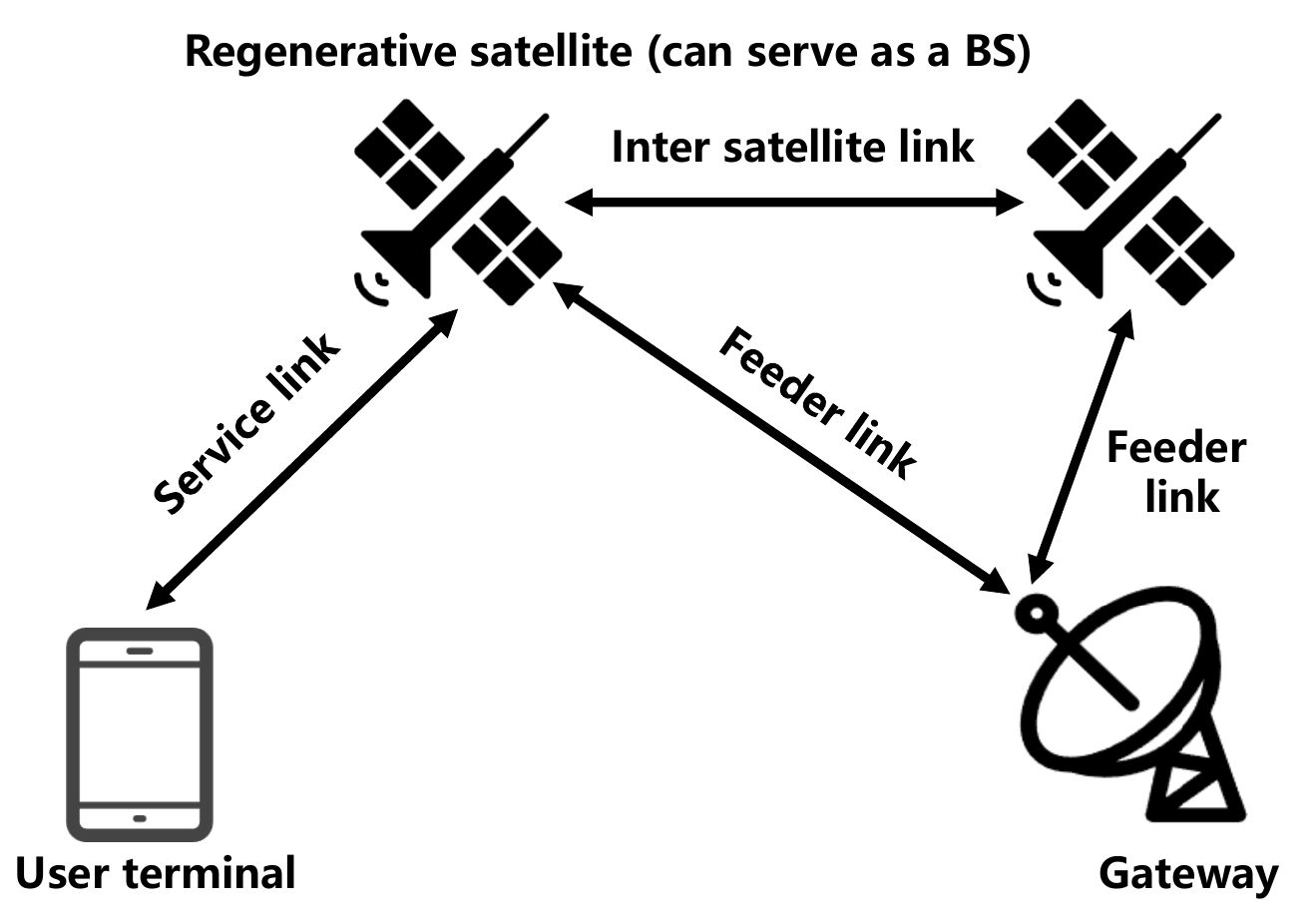}}
	\caption{ SatComs in the regenerative type.}
	\label{RSAT}
\end{figure}

In this subsection, we provide a preliminary background on regenerative SatCom. A satellite has two primary modes of operation: it can either relay signals directly (transparent mode) or perform on-board processing (regenerative mode). The regenerative mode enables a satellite to undertake tasks such as switching, routing, modulation, demodulation, and coding, thereby serving as a BS for the terrestrial network. The architecture of regenerative SatCom types in NTN is illustrated in Fig. \ref{RSAT}, where feeder links connect the gateway to the satellite, and the link between the UT and the satellite is referred to as a service link. In a transparent configuration, on-board processing and inter-satellite links (ISLs) are not supported, necessitating both feeder and service links in SatComs. Conversely, regenerative satellites can utilize ISLs, enabling other satellites to support a service link if a gateway is not within a single satellite's coverage by establishing a feeder link through the gateway.

Satellite communication channels exhibit several significant differences from terrestrial networks. A notable difference is the considerable signal attenuation and long delay caused by the vast distance between the satellite and ground devices. Another critical distinction is the strong LoS channel characteristics, which include path loss, atmospheric attenuation, and the impact of rainfall. Additionally, substantial Doppler shift and CCI are present in satellite channels.

\subsection{Signal and Interference}
In this study, we consider a scenario involving an LEO satellite equipped with a single antenna. The received uplink signal at the satellite is formulated as \cite{you2020massive}
 \begin{equation}
     y^{\rm SAT}(t)=h (t)\star x^{\rm UT} (t)+z (t),
     \label{eq1}
 \end{equation}
 where $\star$ denotes the convolution operation. Here, ${x}^{\rm UT} (t)$ represents the transmit signal at the UT, $h (t)$ the channel impulse response of LEO satellite, and $z (t)$ the additive noise. The channel impulse response is detailed as
 \begin{equation}
     h (t,\tau)=\sum_{l=0}^{L-1}\alpha _l e^{j2\pi v_l t} \delta(\tau-\tau_l),
 \end{equation}
 where $\delta(\cdot)$ indicats the Dirac delta function, $L$ is the number of multipath components, $\alpha _l$ the uplink channel gain, $v_l$ the Doppler shift, and $\tau_l$ the delay for the $l$-th path.
The downlink channel model mirrors the uplink model. Given the satellite's typically higher transmission power compared to UTs, such as smartphones, the downlink channel conditions generally surpass those of the uplink.

SatComs necessitate aggressive frequency reuse due to the vast number of satellites and their movement, leading to a complex CCI environment \cite{nguyen2022outage}. Moreover,  LEOs must avoid spectrum interference with other networks, like geostationary orbit satellites (GEOs), further complicating the scenario. Despite the introduction of spectrum sensing \cite{jia2017joint} and interference mitigation techniques \cite{zhang2023data}, CCI remains a formidable challenge amidst the rapid increase in LEOs and terrestrial devices. Accounting for interference from $N$ devices, (\ref{eq1}) can be reformulated as
  \begin{equation}
     {y}^{\rm SAT}(t)={h} (t)\star {x}^{\rm UT} (t)+\sum_{i=0}^{N-1} {h}^{i }(t)\star {x}^{i}(t)+{z} (t),
 \end{equation}
 where ${h}^{i}(t)$ and ${x}^{i}(t)$ represent the channel response and the transmit signal of other devices, which are unknown to the current transmission.

This study aims to develop a channel-adaptive transmission strategy leveraging generative models, facing the challenges of limited computational resources and training data in satellite communications. Therefore, only compact models are viable for enhancing the capabilities of the FMs. Under such computational constraints, these compact models can improve the foundation models for specific tasks and channel conditions, enabling better adaptation and performance.


\section{Generative FM-based Semantic Transceiver}
\label{s3}

In this section, we delve into the application of FMs for images, focusing on how semantic segmentation models and guided diffusion models address various aspects of semantic image processing, such as the segmentation of specific parts and the repair of damaged areas. Initially, we introduce the concept of a basic semantic transceiver based on FMs. Subsequently, we propose an adaptive semantic satellite transmission architecture that takes into account both user requirements and channel conditions. Furthermore, we explore the utilization of previously well-transmitted images for the restoration of images disrupted by sudden interference. Finally, we present a novel error detection method designed to meet user requirements while optimizing the balance between transmission delay and computational resources.

\subsection{Basic Architecture}
\label{s3a}

\begin{figure*}[h]
	\centering

  {

  \includegraphics[width=0.8\linewidth]{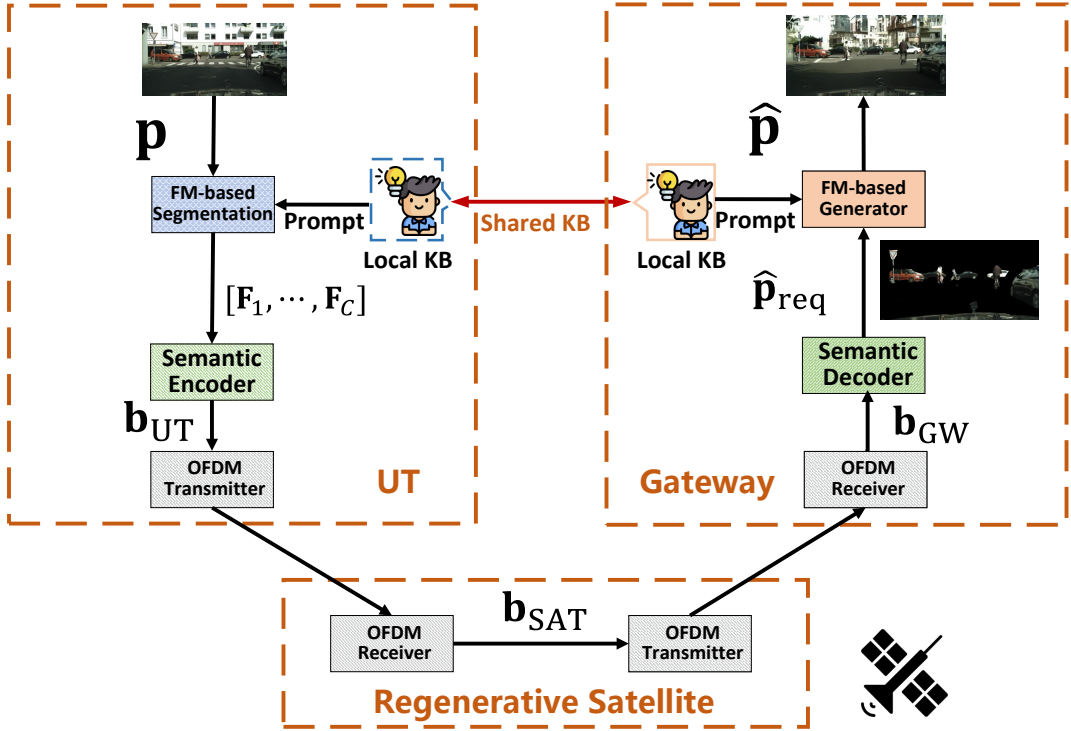}}
	\caption{Backbone framework of FMSAT.}
	\label{Basic_overall}
\end{figure*}

Fig. \ref{Basic_overall} depicts the backbone framework of FMSAT. The source image $\mathbf{p} $ is processed by a FM to segment all semantic features according to user requirements, denoted as prompt $\boldsymbol{\Omega}_{\rm TX}$. These semantic features are represented as
\begin{equation}
    [\mathbf{F}_{1},\ldots, \mathbf{F}_{C}]={\tt SC}_{\rm seg}(\mathbf{p},\boldsymbol{\Omega}_{\rm TX}),
\end{equation}
where ${\tt SC}_{\rm seg}(\cdot)$ symbolizes FM-based semantic segmentation, dividing the input image into $C$ categories.

Subsequently, a CNN-based semantic encoder is utilized to transmit and receive these semantic features, symbolized as ${\tt SC}_{\rm en}(\cdot)$ and ${\tt SC}_{\rm de}(\cdot)$. Each output value of the semantic encoder is one-bit quantified, and these bits are the transmitted codeword, which can be written as
\begin{equation}
    \mathbf{b}_{\rm UT}={\tt SC}_{\rm en}([\mathbf{F}_{1},\ldots, \mathbf{F}_{C}]; \mathbf{W}_{\rm en}),
\end{equation}
where $\mathbf{W}_{\rm en}$ denotes the set of trainable parameters in the semantic encoder, and $\mathbf{b}_{\rm UT}$ is the codeword at the UT. The codeword $\mathbf{b}_{\rm UT}$ is then modulated into $x^{\rm UT}(t)$ through an OFDM transmitter and sent to a regenerative satellite according to Eq. (3). This satellite processes the received signal $y^{\rm SAT}(t)$ akin to a ground BS. Trough the OFDM receiver in the satellite, the codeword can be regenerated in the satellite, denoted as ${\mathbf{b}}_{\rm SAT}$. The HARQ and the ISL are employed to enhance performance and extend coverage. From the satellite to the gateway, the codeword ${\mathbf{b}}_{\rm SAT}$ is similarly sent to the gateway by an OFDM transceiver, where the received codeword is called ${\mathbf{b}}_{\rm GW}$. In semantic communications, transmission errors are somewhat tolerable provided that task-related key features remain intact. Employing a semantic decoder to correct errors in the satellite is not advisable due to the significant increase in transmission delay it causes. The differences between conventional and semantic methods through the regenerative satellite are as follows:

1) The conventional cyclic redundancy check (CRC) error detector ensures ${\mathbf{b}}_{\rm SAT}=\mathbf{b}_{\rm UT}$ via conventional satellite transmission.

2) Errors from radio noise and CCI accumulate may completely distort semantic features, necessitating a novel error detector for semantic errors.

Generative diffusion models operate by gradually transforming a sample from a simple distribution (often Gaussian noise) into a sample from the target distribution through a Markov process. The image of $T$-th timestep $\mathbf{p}_{T}$ is equal to a fully noised sample $\mathbf{z}$. A neural network is trained to convert $\mathbf{p}_{t}$ into $\mathbf{p}_{t-1}$ and finally $\mathbf{p}_{0}$ is equal to the target image $\hat{\mathbf{p}}$. This network can be represented by a conditional probability $p(\mathbf{p}_{t-1}|\mathbf{p}_{t})$. Moreover, a condition at the receiver $\boldsymbol{\Omega}_{\rm RX}$ can be added to control the generated image while the neural network learns $p(\mathbf{p}_{t-1}|\mathbf{p}_{t},\boldsymbol{\Omega}_{\rm RX})$. The entire procedure is represented as 
\begin{equation}
    \hat{\mathbf{p}}={\tt SC}_{\rm difu}(\mathbf{z},\boldsymbol{\Omega}_{\rm RX}),
\end{equation}
where the inputted condition $\boldsymbol{\Omega}_{\rm RX}$ is consistent of received semantic features and other prompts at the gateway, yielding
\begin{equation}
\boldsymbol{\Omega}_{\rm RX}=[{\tt SC}_{\rm de}({{\mathbf{b}}_{\rm GW}} ),\ldots].
\end{equation}
A straightforward approach combines state-of-the-art methods such as SegGPT and the conditional diffusion model for segmentation and reconstruction. The semantic encoder-decoder is trained to convert these semantic features into the codeword. This transmission process using SegGPT is referred to as FMSAT(SegGPT). Thanks to the generic capabilities of these FMs, FMSAT outperforms conventional semantic communications, like joint source-channel coding (JSCC). However, this architecture may not be suitable for SatComs under limited complexity and unfamiliar transmission noise conditions.

The above semantic framework can work well by directly combining these foundation models with the current semantic communication strategies. However, the characteristics of the communication devices and channels should also be fully exploited and two changes are made as follow.

1) \textbf{Reducing complexity at the UT:} SegGPT is a powerful semantic segmentation method for widely usage. Considering the limited resource at the UT,
a practical method is utilizing SegGPT to teach a simpler UNet for semantic segmentation. Especially when labeled data are scarce, segmented images from SegGPT serve as references, where the mean Intersection over Union (mIoU) performance of SegGPT is close to 80\%, but that of UNet falls to nearly 60\% under such conditions. As a student, the simple UNet under insufficient training data is weaker but still acceptable since detailed parts are more error-prone than major ones. This phenomena is beneficial for semantic transmission because the important part of an image usually occupies a major space. The transmission method utilizing UNet for segmentation is termed FMSAT(UNet).

2) \textbf{Enhancing performance at the gateway:} The pretrained diffusion model can be used as a roust method for denoising and inpainting. To adapt to the communication system, the conditional diffusion model is trained to generate the target images using the inputted conditions with transmission errors.

\subsection{Adaption and Cooperation}

\begin{figure*}[h]
	\centering

  {

  \includegraphics[width=0.99\linewidth]{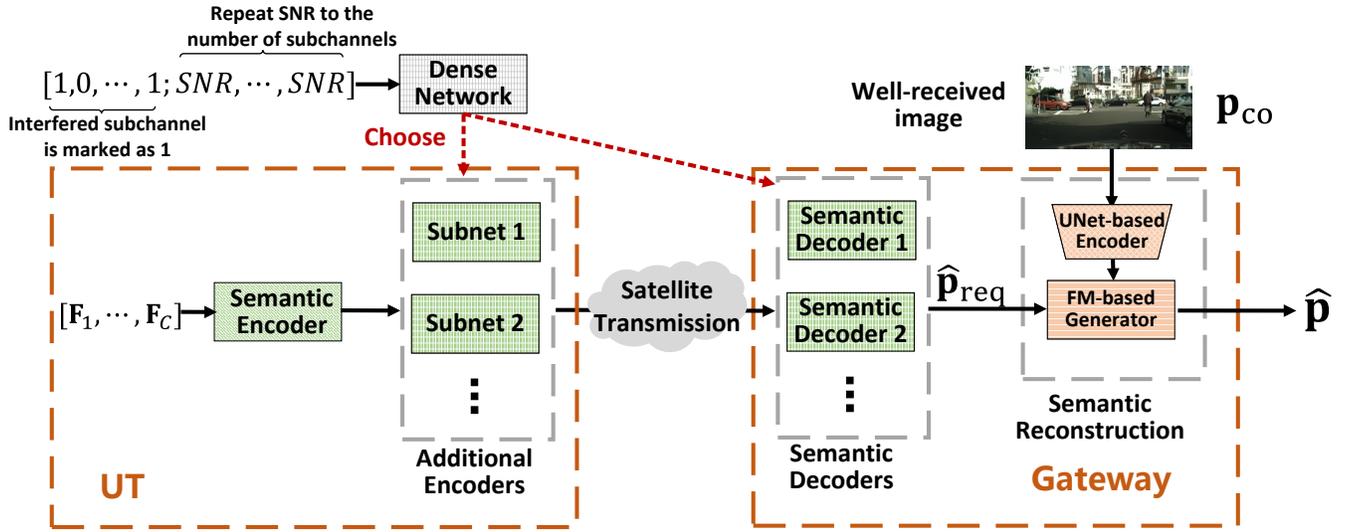}}
	\caption{ Framework of adaptive FMSAT with correlation. The different additional encoder and decoder pairs are trained under different channel conditions. }
	\label{AFMSAT_overall}
\end{figure*}

To further enhance transmission performance, adaptive methods are employed, leveraging advancements in CSI feedback and interference prediction \cite{lu2018new}. Assuming the availability of CSI at the transmitter, the transmitted parts are selectively chosen for optimal reconstruction. A multi-rate semantic encoder-decoder is utilized, incorporating a set of additional encoder at the transmitter to manage the transmitted information effectively. Different semantic decoders at the gateway corresponding to the specific encoder are required.

As depicted in Fig. \ref{AFMSAT_overall}, some additional encoders $f_{{\rm en},i}()$  following the semantic encoder generates the codeword $\mathbf{b}_{\rm UT}$ based on the channel condition. The semantic encoder uses the well-trained parameters $\hat{\mathbf{W}}_{{\rm en}}$ from Section III A and these parameters are fixed. These additional encoders are simple due to the limited resource at the UT. The output dimension of these additional encoders shall be adjusted based on the available bandwidth.
At the gateway, the entire semantic decoder is trained for the additional encoder to reach a good performance. Thus, all the additional encoders have their corresponding semantic decoders.  
The loss function is redesigned to guide the additional encoders and semantic decoders in restoring required image part $\hat{\mathbf{p}}_{\rm req}$ under a specific channel condition. The process is formulated as
\begin{equation}
\begin{aligned}
 &\hat{\mathbf{p}}_{\rm req}=\\&{\tt SC}_{\rm de}(h^{\dag}(f_{{\rm en},i}({\tt SC}_{\rm en}([\mathbf{F}_1,\ldots,\mathbf{F}_{C}];\hat{\mathbf{W}}_{{\rm en}});\mathbf{W}_{{\rm en},i}));\mathbf{W}_{{\rm de},i}),  
\end{aligned}
\end{equation}
where $\mathbf{W}_{{\rm en},i}$ and  $\mathbf{W}_{{\rm de},i}$ represent the trainable parameters in the $i$-th encoder and $i$-th decoder, respectively. Due to the practical channel $h(t)$ is not suitable for end-to-end training, the channel impact $h^{\dag}(\cdot)$ is simulated by adding noise and masking the interfered part. Several pairs of encoders and decoders are trained for different scenarios. For example: 
\begin{itemize}  
    \item With no interference and SNR = 10 dB, the entire image is required to be well-transmitted at this good channel condition. The encoder-decoder is trained by minimizing the mean squared error (MSE) of each pixel. %
    
    \item With no interference and SNR = -10 dB, only important parts are well-transmitted. These parts are chosen according to user requirement. The encoder-decoder is trained to restore these image parts. The MSE loss is only calculated on the pixel at the required parts while errors at the other pixels are ignored, which is called required MSE.  
    \item With 50\% interference and SNR = 0 dB, the encoder's output is halved  and transmitted over interference-free subchannels. Meanwhile,  only important parts are well-transmitted. The loss function targets the required MSE.
\end{itemize}
Settings can be adaptively modified to optimize performance, similar to selecting code rates in conventional methods. Upon identifying the current channel condition, a simple dense network is trained to select the appropriate additional encoder and decoder pair for enhanced performance. The conditions of interfered subchannels and the average SNR are inputted for choice. The single value of the average SNR is repeated across all subchannels to capture the network's attention.


Context correlation presents a promising avenue to improve performance, especially in high-mobility SatComs. Given the rapid changes in channel conditions, an image may deteriorate suddenly, while some well-received images remain accessible in the same region. Although a well-received image from different shooting angles and locations might lack certain important information, like vehicles, useful semantic features such as architectural style and weather conditions persist. This indicates that shared features in the region can always aid in restoring a compromised image. The generation of a better image through the cooperation of a correlated good image $\mathbf{p}_{\rm co}$ and a poor current image exemplifies the strength of an FM-based architecture. The entire process is described as
\begin{equation}
     \hat{\mathbf{p}}={\tt SC}_{\rm difu}(\mathbf{z},[\hat{\mathbf{p}}_{\rm req}, \mathbf{p}_{\rm co}]).
\end{equation}

\subsection{Novel Error Detector for FMSAT}

\begin{figure*}[h]
	\centering

  {

  \includegraphics[width=0.99\linewidth]{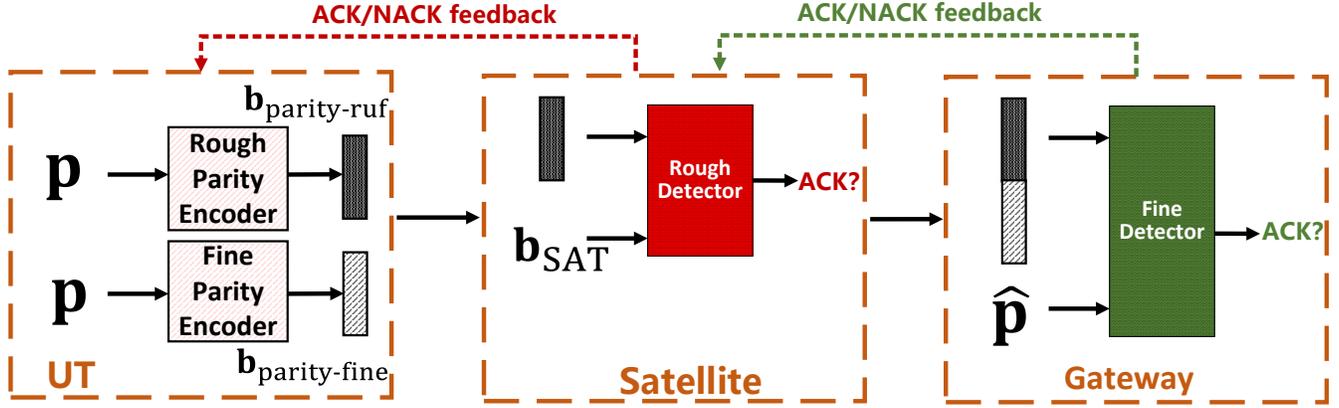}}
	\caption{The proposed error detectors at the satellite and the gateway.}
	\label{Err_overall}
\end{figure*}

According to the HARQ process \cite{3GPP}, retransmission operations depend on the feedback, which should be sent to the transmitter before the deadline. For long propagation delay, preemptive re-transmission without ACK is a good choice. In this situation, the receiver could send ACK after combining the received packets. However, this HARQ mode requires more transmission bandwidth at the first transmission for repeatedly sending. Overall, obtaining feedback as soon as possible is always beneficial to the system design. The regenerative satellites rely on signal processing at the satellite. For HARQ techniques in SatComs, transmission delays in regenerative satellite can be approximated as half of those in the transparent one because the error detector can be used in the satellite.

Owing to the reconstruction capability of the FM modules, some transmission errors have no influence on the received image. That means directly using the CRC error detector requires unnecessary retransmission. The existing image assessors can detect the quality of received ones but have two drawbacks when applied in the regenerative SatComs: 1) Reconstructing the image at the satellite is not a good choice due to the limited computational resources, and this process delays ACK feedback; 2) The required parts should be real rather than acceptable. Thus, a two-level error detector is proposed to solve these issues as shown in Fig. \ref{Err_overall}.

The first error detector, called the rough detector, is used at the satellite and aims at detecting the errors in the required parts quickly. The input of the rough detector is a 32-bit parity code, $\mathbf{b}_{\rm parity-ruf}$, and received codeword ${\mathbf{b}_{\rm SAT}}$. The parity code for rough detection can be generated by a CNN-based encoder $f_{\rm ruf}(\cdot)$ as
\begin{equation}
\mathbf{b}_{\rm parity-ruf}=f_{\rm ruf}(\mathbf{p}).
\end{equation}
The output of the rough detector is one value with a sigmoid activation function. A hard decision converts this output into 0 or 1. If the output value is 0, then the received codeword is good and can be transmitted to the gateway. Otherwise, the NACK is sent to the transmitter, and retransmission is required. This process can be expressed as 
\begin{equation} 
    \left\{ \begin{aligned} 
        &{\rm ACK},&{\rm if} ~ D_{\rm ruf}(\mathbf{b}_{\rm parity-ruf},{\mathbf{b}_{\rm SAT}})>0.5,\\
       &{\rm NACK}, & ~ {\rm otherwise},
    \end{aligned}\right. \label{eq10}
\end{equation}
where $D_{\rm ruf}(\cdot)$ is the rough error detector at the satellite.

The key step is to train the rough detector. The training data is generated by transmitting the images under different channel conditions and calculating the MSE performance at the required parts between the transmit and received images. Thus, the label can be written as
\begin{equation}
   {\rm Label} = \left\{ \begin{aligned}
        &1,&{\rm if} \quad {\rm MSE}_{\rm req} ({\tt SC}_{\rm de}({\mathbf{b}_{\rm SAT}}),\mathbf{p} ) \leq 0.015 ,\\
       &0, & {\rm otherwise},
    \end{aligned}\right. 
\end{equation}
where ${\rm MSE}_{\rm req}$ only calculates the MSE between the pixels of the required parts according to the task requirement. The threshold is set to be 0.015 to ensure that the required parts are not destroyed by transmission errors. The simple rough detector may also make a mistake when recognizing the semantic loss in the required parts from the transmit codeword and the parity code. Thus, a loose threshold is a good choice to avoid improperly sending NACK for a well-received codeword.

The second error detector, called the fine detector, is used at the gateway, where the image has been reconstructed. The parity code $\mathbf{b}_{\rm parity-fine}$ is also generated by the UT and the ACK/NACK is generated by the gateway similar to Eq. (\ref{eq10}), while the threshold is calculated by combining the required MSE, ${\rm MSE}_{\rm req}(\cdot) $, and a pre-trained image quality assessment (ASS) \cite{idealods2018imagequalityassessment}, yielding
  \begin{equation}
   {\rm Label} = \left\{ \begin{aligned}
        &1,&{\rm if} \quad {\rm MSE}_{\rm req} (\hat{\mathbf{p}},\mathbf{p} ) \leq 0.015 \quad {\rm and} \quad {\rm ASS}>4.5,\\
       &0, & {\rm otherwise}.
    \end{aligned}\right.
\end{equation}

In summary, FM-based semantic communications can potentially improve the transmission efficiency of SatComs. The channel characteristics and the transmission framework with regenerative satellites are considered in this study. The FM-based models are fully utilized to enhance the performance at the UT, satellites, and the gateway.

\section{Architecture And Training Process }
In this section, the models employed within the FMSAT will be introduced in detail. 
Following the introduction of these models, the training process of each component is meticulously described.

\subsection{Detailed Network Architectures}

%
\begin{figure*}[t]
	\centering

		{\includegraphics[width=0.99\linewidth]{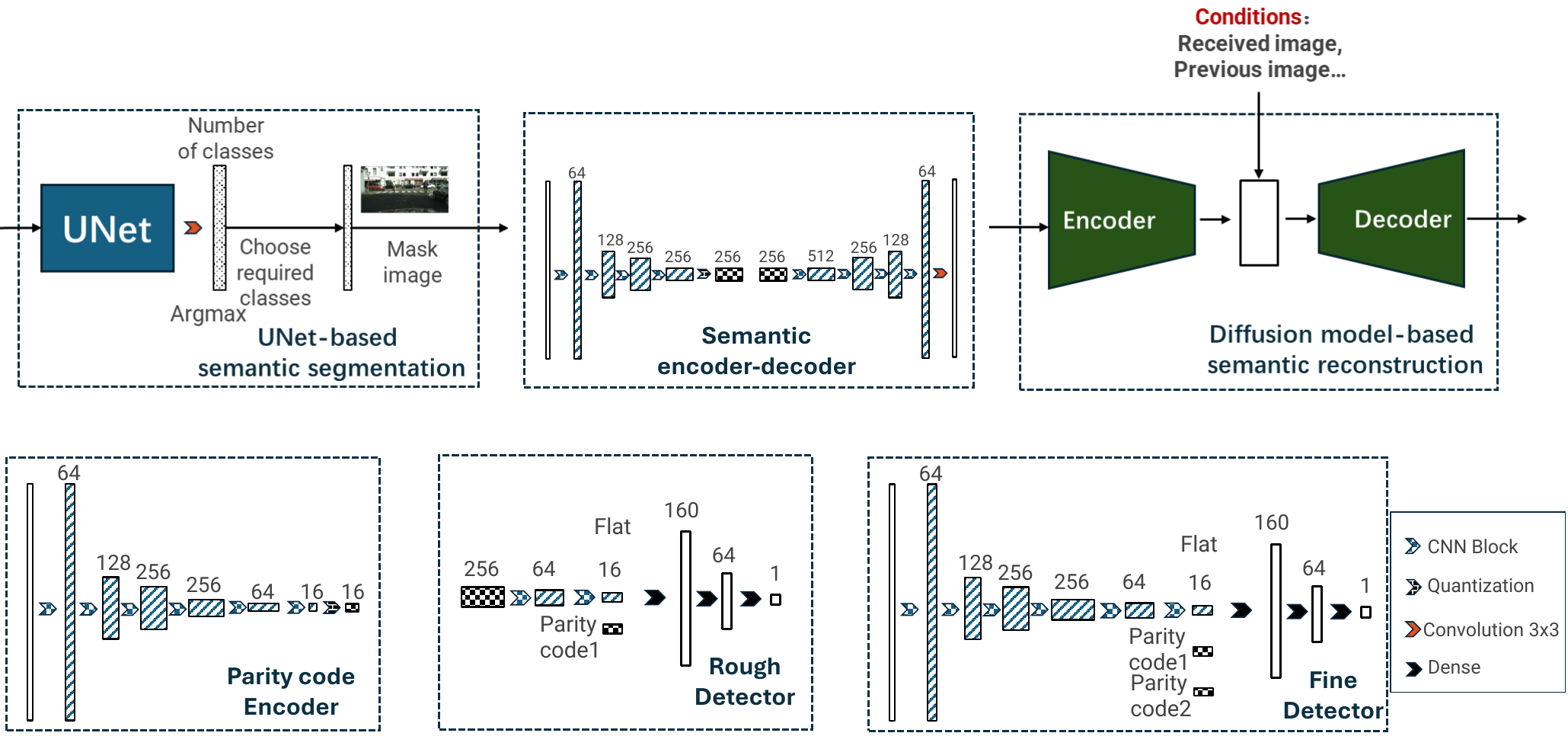}}

	\caption{Architectures of the semantic segmentation and encoder-decoder. The number of output channels  is marked above each CNN-based operation. }
	\label{Detailed modules}
\end{figure*}

The architecture of FMSAT is designed to facilitate efficient semantic communication for satellite systems. As shown on the left of Fig. \ref{Detailed modules}, it consists of several key components, each tailored to enhance the transmission of semantic information under varying channel conditions. The main components include:

\textbf{FM-based semantic segmentation ${\tt SC }_{\rm seg}(\cdot)$:} 
As presented on the left side of Fig. \ref{Detailed modules}, the study examines two semantic segmentation approaches. SegGPT, known for its excellent performance across various scenarios and requirements, achieves a mIoU near 80\%. In scenarios where training labels for changing transmission content are unavailable, a UNet-based compact semantic segmentation approach is employed, trained with labels generated by SegGPT. This UNet structure comprises four CNN blocks in both encoder and decoder sections, with each block containing two convolution operations followed by 2 $\times$ up/downsampling. The encoder's hidden channels sequence is 64, 128, 256, and 512, while the decoder's sequence is 1024, 512, 256, and 128. To align the decoder's output with the required categories, two final convolution operations adjust the output to 64 channels.


\textbf{Semantic encoder and decoder ${\tt SC}_{\rm en}(\cdot)$,  ${\tt SC}_{\rm de}(\cdot)$:} Detailed in Fig. \ref{Detailed modules}, the encoder features four convolution operations with window sizes of $9\times 9$,  $7\times 7$, $5\times 5$, and $3\times 3$ respectively, and hidden channels of 64, 128, 256, and 256. The stride for the first convolution is set at 4, with subsequent convolutions at 2, incorporating the ReLU activation function. The dimension of encoder's output is (256, 16, 8) and a quantization layer converts this output into binary form. The decoder mirrors the encoder's process with reversed channels and concludes with a $3\times 3$ convolution operation utilizing a Tanh activation function, producing an output image of dimensions (512, 256, 3).

\textbf{Additional encoder and corresponding decoder $f_{{\rm en},i}(\cdot)$, ${\tt SC}_{\rm de}(\cdot)$:} The additional encoder includes two $3 \times 3$ convolution operations with 512 and $K$ channels---$K$ being determined by the available transmission bandwidth. The initial layer employs ReLU activation, with the final layer using Tanh alongside quantization. The corresponding decoder mirrors the basic decoder but adjusts its input channels to match the adaptive encoder's output.


\textbf{Error detectors:} A CNN-based encoder generates a 32-bit parity code, with the initial three convolutions similar to the semantic encoder with quantization operation. The subsequent two $3 \times 3$ convolutions have strides of two and channels of 64 and 16, respectively. Thus, the output dimension of the parity code encoder is (16, 2, 1). Post-quantization, this encoder produces a 32-bit parity code. The rough detector compresses the codeword from (256, 16, 8) to (16, 4, 2) dimensions through two convolutions with strides of two at the satellite, integrating it with the parity code before employing two dense layers for ACK/NACK decision output. The fine detector, positioned at the gateway, similarly compresses the received image to (16, 4, 2), while merging these compressed features with two parity codes for final decision-making.

\textbf{FM-based semantic reconstruction ${\tt SC}_{\rm difu}(\cdot)$:} A conditional diffusion model is utilized for denoising and repairing received images. Conditions encoded from the received image via a UNet-based encoder extract vital information from damaged images, highlighting the reconstruction process's adaptability and effectiveness in ensuring the integrity of transmitted data.

\subsection{Training Process}

\textbf{Semantic segmentation:}
To learn the segmentation results from SegGPT, the segmented results are saved. The prompt is tailored to the dataset, identifying objects such as vehicles, signs, and buildings in city images. Subsequently, the UNet is trained using these results as labels. The cross-entropy function serves as the loss function, with Adam being the optimizer at a learning rate of 0.0005.

\begin{figure*}[h]
	\centering

		{\includegraphics[width=0.99\linewidth]{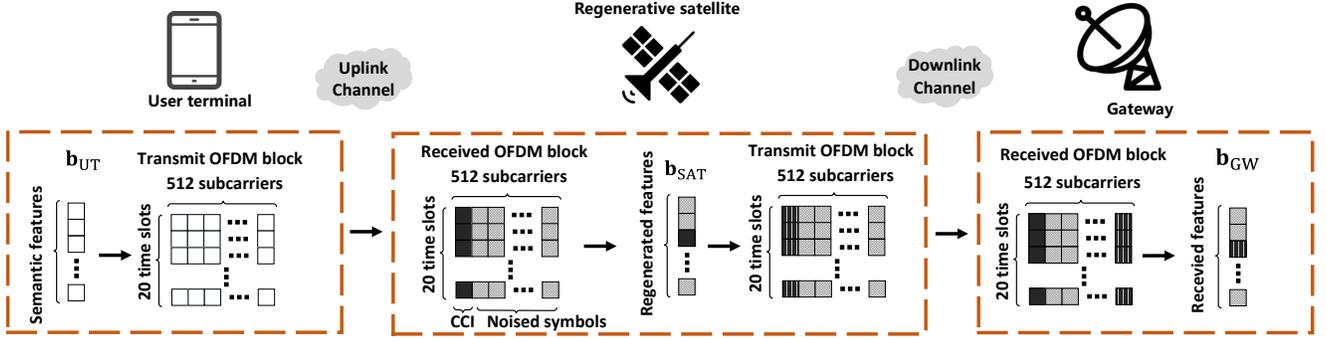}}

	\caption{The transmission frame of the regenerative SatCom. }
	\label{Frame}
\end{figure*}

\textbf{Semantic encoder-decoder and adaptive methods:}
The basic semantic encoder-decoder discussed in Section III A undergoes end-to-end training at an SNR of 10 dB without CCI. The MSE is the chosen loss function, and the training process is formulated as
\begin{equation}
    \widehat{\mathbf{W}}_{{\rm en}},\widehat{\mathbf{W}}_{{\rm de}}=\mathop{\arg\min}\limits_{\mathbf{W}_{{\rm en}},\mathbf{W}_{{\rm de}}} \left\|\mathbf{p}-{\tt SC}_{\rm de}(h^{\dag}({\tt SC}_{\rm en}({\tt SC}_{\rm seg}(\mathbf{p})))) \right\|^2.
\end{equation}

For training adaptive additional encoders and decoders, the trained weights are utilized as a starting point. The different channel conditions are represented as $h_1^{\dag}(\cdot)$, $h_2^{\dag}(\cdot)$, and $h_3^{\dag}(\cdot)$, trained alternately within an epoch. A mask $\Theta$, derived from the segmented results, directs the encoder-decoder to concentrate on significant parts. During training under the channel condition $h_i(\cdot)$, the training process is
\begin{multline}  
     \widehat{\mathbf{W}}_{{\rm en}},\widehat{\mathbf{W}}_{{\rm en},i},\widehat{\mathbf{W}}_{{\rm de},i}=\\  \mathop{\arg\min}\limits_{\mathbf{W}_{{\rm en}},\mathbf{W}_{{\rm en},i},\mathbf{W}_{{\rm de},i}} {\left\| \Theta \odot \mathbf{p}- \Theta \odot {\tt SC}_{\rm de}(h_i^{\dag}({\tt SC}_{\rm en}({\tt SC}_{\rm seg}(\mathbf{p})))) \right\|}^2,
\end{multline} 
where $\odot $ denotes element-wise product.
Upon completion of training, these additional encoder and decoder pairs are equipped to operate under diverse conditions. A straightforward dense network is then trained to select the specific pair for the current condition, with the channel condition represented as $$[1,1,0,\ldots,1, {\rm SNR},\ldots, {\rm SNR}]$$ In this representation, '1'  indicates subcarriers disrupted by CCI, while '0' marks unaffected subcarriers. The sequence repeating the current SNR corresponds to the number of subcarriers to attract the network's attention. The hidden layer, consisting of 512 neurons with the ReLU activation function, leads to a final layer of three neurons with the Softmax activation function. The labels are generated by selecting the additional encoder and decoder pair that minimizes the MSE of crucial semantic features under the present channel conditions.

\textbf{Semantic reconstruction:}
A widely acknowledged conditional diffusion model is employed for the reconstruction of received images. The conditions are based on the semantic features received post-decoding, with the original image serving as the label. Training the conditional diffusion model across varied transmission outcomes proves advantageous, not only for denoising or inpainting but also for leveraging information within the received semantic features to produce high-quality images. Moreover, incorporating an adjacent well-received image as a condition further enhances the image reconstruction process.

\textbf{Error detector:}
The error detector adopts an end-to-end structure across different conditions. Results obtained using the above transmission framework are collected, with the label of ACK/NACK determined by the proposed threshold. The loss function employed is cross-entropy $L_{\rm CE}(\cdot)$, and the rough detector's training process is described as:
\begin{equation} 
    \widehat{\mathbf{W}}_{D_{\rm ruf}},\widehat{\mathbf{W}}_{{f_{\rm ruf}}}= 
    \mathop{\arg\min}\limits_{\mathbf{W}_{{D_{\rm ruf}}},\mathbf{W}_{{f_{\rm ruf}}}} L_{\rm CE} {\left( {\rm Label} , D_{\rm ruf}(\mathbf{b}_{\rm SAT},f_{\rm ruf}(\mathbf{p})  ) \right)},
\end{equation}
where $\mathbf{W}_{{D_{\rm ruf}}}$ and $\mathbf{W}_{{f_{\rm ruf}}}$ denote the trainable parameter sets for the rough detector and the parity code encoder, respectively. The training methodology for the fine detector mirrors that of the rough detector.

\section{Simulations Results}
\label{s5}

In this section, we conduct several simulations to demonstrate the impact of FMSAT in SatComs. Initially, we detail the simulation settings.We then compare FMSAT against competing methods under various channel conditions, illustrating the effectiveness of our proposed adaptation and correlation strategies. Finally, we underscore the benefits of our proposed error detector in striking an optimal balance between transmission delay and the quality of the received images. 

\subsection{Settings}

\textbf{Satellite scenarios:} We utilize a NTN-TDL-A channel model. The LEO satellite operates at a speed of 7.5622 km/s and an altitude of 600 km, while a mobile UT moves at 3 km/h. Images are transmitted from the UT to the regenerative satellite via an uplink channel, employing a OFDM transceiver. Thus, the transmit codeword is regenerated by an OFDM receiver in the satellite. These bits are then verified and forwarded to the gateway. As depicted in Fig. \ref{Frame}, a frame encompasses 20 time slots, each with 512 subcarriers. Sixteen time slots transmit 16-QAM modulated image bits, while the remaining four slots serve as pilots. The propagation delay of the LEO satellite varies between 2 and 20 ms as it orbits.

\begin{figure*}[h]
	\centering
	\subfigure[]{
		
		{\includegraphics[width=0.45\linewidth]{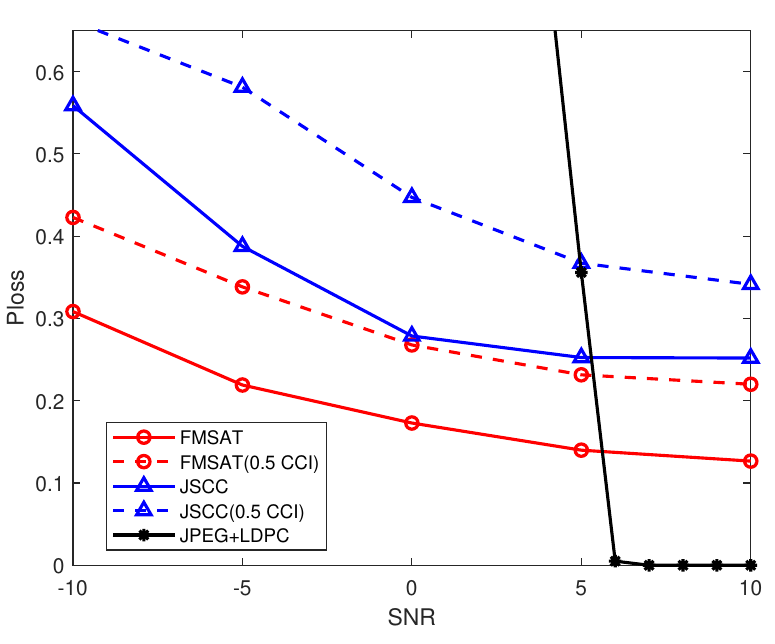}}}
\subfigure[]{
		
		{\includegraphics[width=0.45\linewidth]{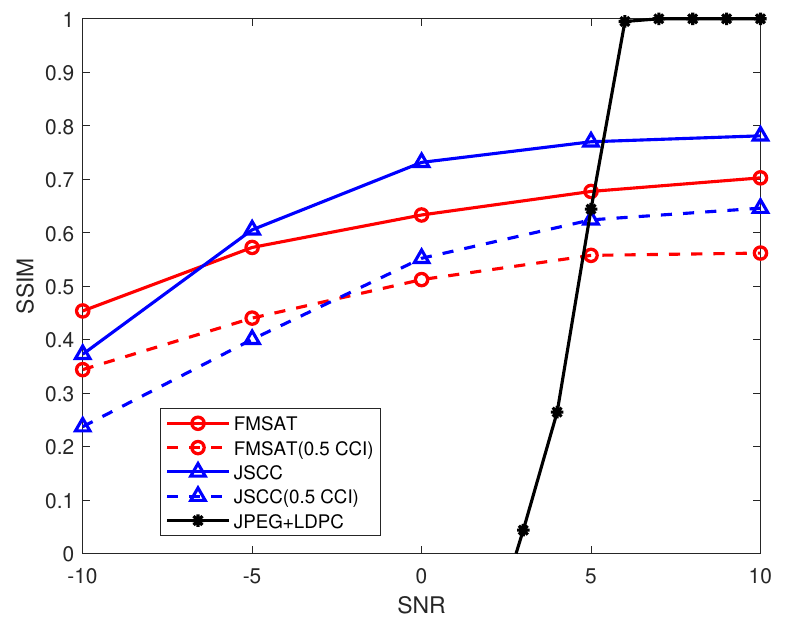}}}
	\caption{Performances of the competing methods under different channel conditions. (a) Ploss performance.  (b) SSIM performance.}
	\label{Basic_Per}
\end{figure*}

\begin{figure*}[h]
	\subfigure[FMSAT, 0 CCI, 0dB SNR]{
		{\includegraphics[width=0.235\linewidth]{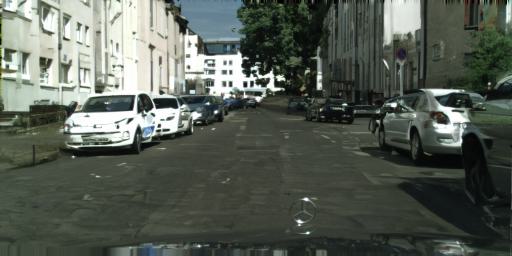}}}
  \subfigure[JSCC, 0 CCI, 0dB SNR]{
		{\includegraphics[width=0.235\linewidth]{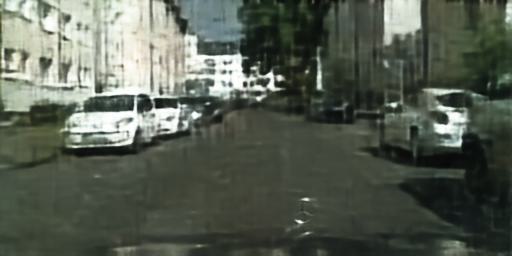}}}
\subfigure[FMSAT, 0 CCI, -10dB SNR]{
		{\includegraphics[width=0.235\linewidth]{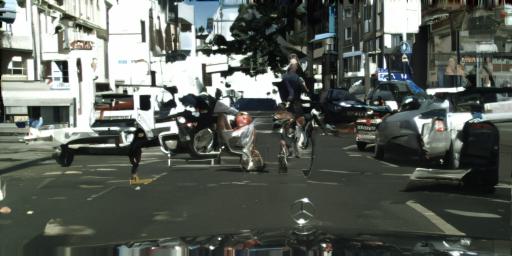}}}
\subfigure[JSCC, 0 CCI, -10dB SNR]{
		{\includegraphics[width=0.235\linewidth]{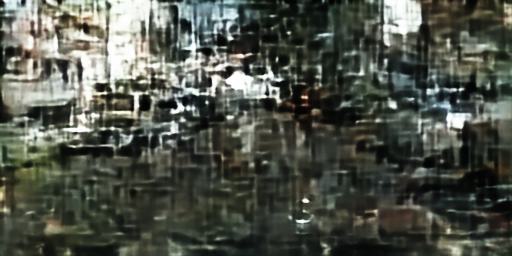}}}\\
  \subfigure[FMSAT, 0.5 CCI, 0dB SNR]{
		{\includegraphics[width=0.235\linewidth]{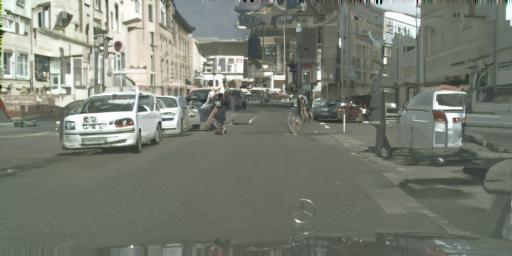}}}
  \subfigure[JSCC, 0.5 CCI, 0dB SNR]{
		{\includegraphics[width=0.235\linewidth]{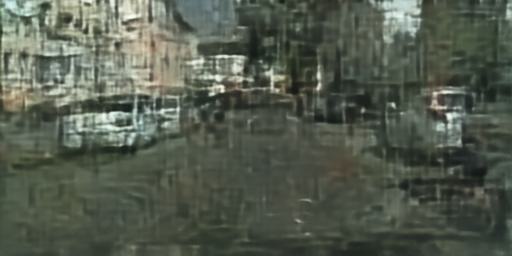}}}
\subfigure[FMSAT, 0.5 CCI, -10dB SNR]{
		{\includegraphics[width=0.235\linewidth]{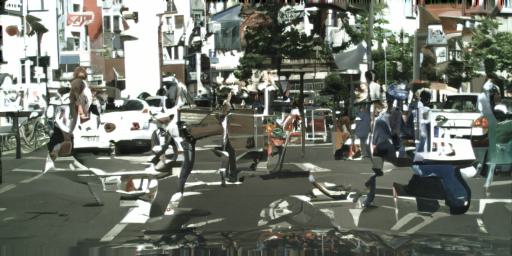}}}
\subfigure[JSCC, 0.5 CCI, -10dB SNR]{
		{\includegraphics[width=0.235\linewidth]{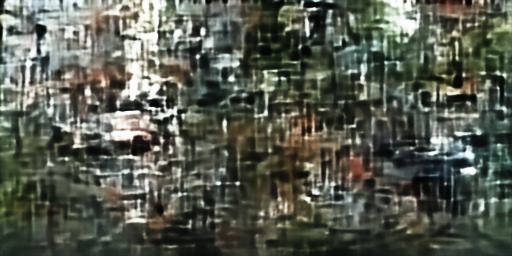}}}\\

	\caption{Image examples restored by the competing methods under different transmission scenarios.}
	\label{Basic_example}
\end{figure*}

In addition to experiencing dramatically fluctuating SNRs across different frames, CCI is also simulated. Some subchannels are disrupted by signals from other devices. For simplicity, the transmit signal in these subchannels is subjected to significant noise in this study, representing the most adverse condition. It is assumed that the transceiver is aware of the affected subchannels when interference originates from predictable sources, such as other satellites. However, interference from numerous ground devices is unpredictable and is considered to occur randomly.

\textbf{Dataset:} The Cityscapes and COCO-stuff datasets are employed to test the basic FMSAT. Additionally, a well-received cityscape image from the same city is used for the correlation method. All images are resized to (512, 256, 3). The training dataset is tenfold larger than the testing set. 

\textbf{Metrics:} (1) \textbf{Perceptual loss (Ploss)}: Ploss is widely used to measure the feature discrepancy between estimated and actual images. A pre-trained network like VGG is used to quantify feature similarity across various layers, as denoted by Ploss \cite{zhang2018unreasonable}. (2) \textbf{Structural similarity index (SSIM)}: SSIM evaluates the structural resemblance among image patches \cite{hore2010image}. 

\textbf{Competing method:} The conventional technique involves using JPEG for source coding and Low-Density Parity-Check (LDPC) codes for channel coding to transmit an image. This method results in an image size of approximately 144 Kbits after JPEG compression, with an LDPC code rate of 64/127. Conversely, semantic methods require only 32 Kbits for image transmission. Moreover, the JPEG+LDPC approach employs 4-QAM modulation for efficient performance under low SNRs but necessitates nearly 18 times more OFDM symbols compared to semantic approaches. Additionally, JSCC, a classic semantic method, uses a CNN-based encoder-decoder.

\subsection{Performances of Basic FMSAT}

In this subsection, we evaluate the performance of the proposed methods under various channel conditions, excluding the impact of the error detector. Therefore, transmission errors from the UT-satellite and satellite-gateway links are accumulated. The SNR tested reflects the total SNR experienced by the transmit features through the SatCom system. Additionally, \textbf{0.5 CCI} implies random interference at subchannels, with broken semantic features occupying, on average, 50 percent of a received codeword.

Fig. \ref{Basic_Per} demonstrates the effectiveness of the proposed methods across different channel conditions. The conventional JPEG+LDPC method excels at high SNRs but sees a marked decline in performance as transmission errors surpass its error correction capability, indicating that traditional SatCom methods demand additional resources to maintain SNR above a certain threshold. In Fig. \ref{Basic_Per}(a), the FMSAT method consistently outperforms the JSCC approach in Ploss, suggesting that semantic segmentation and reconstruction via generative FMs significantly enhance received image quality. SSIM results, shown in Fig. \ref{Basic_Per}(b), reveal that at high SNR, the images processed by JSCC are closer to the original compared to those by FMSAT. However, under deteriorating channel conditions, especially with interference, FMSAT's SSIM outstrips JSCC's, indicating FMSAT's superior capability in image restoration from fragmented features, a feat not as effectively achieved by JSCC.


Visual comparisons in Fig. \ref{Basic_example} starkly illustrate the differences between these methods. Images processed by JSCC become increasingly blurred with noise and interference augmentation, becoming entirely unrecognizable under 0.5 CCI and -10 dB SNR. In contrast, FMSAT maintains image clarity, although some objects may alter due to transmission feature disruptions. These alterations become more pronounced under CCI conditions. For instance, FMSAT images under 0.5 CCI and 0 dB SNR exhibit altered brightness, and those under 0.5 CCI and -10 dB SNR display non-existent objects, reflecting ineffective information for reconstruction due to received feature impairment. Furthermore, high noise levels distort object outlines in images processed by FMSAT at -10 dB SNR.

\begin{figure}[h]
	\centering
	\subfigure[]{
		\label{Diff_RIS_ACC}
		{\includegraphics[width=0.9\linewidth]{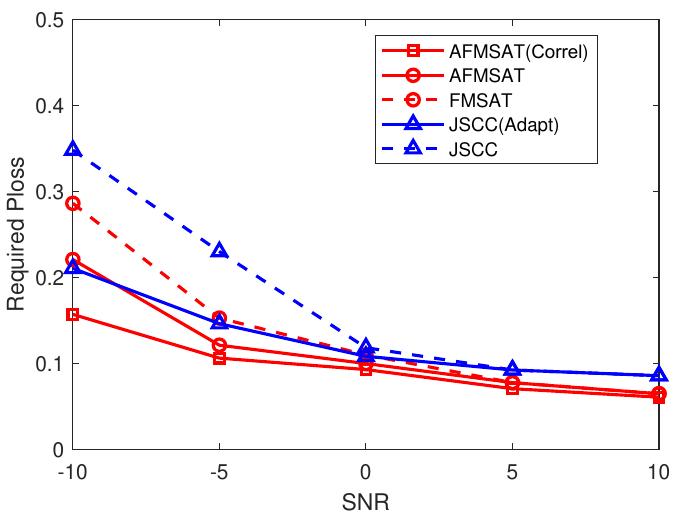}}}\\
\subfigure[]{
		\label{Diff_RIS_MSE}
		{\includegraphics[width=0.9\linewidth]{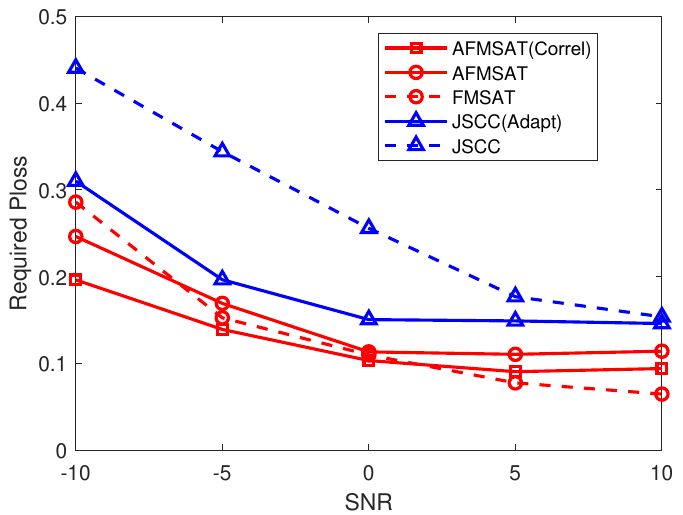}}}
	\caption{ Ploss performance of the required semantic feature restored by the proposed methods.}
	\label{ADA_Per}
\end{figure}

 \begin{figure}[h]
	\flushleft
  \subfigure[Transmit]{
		{\includegraphics[width=0.45\linewidth]{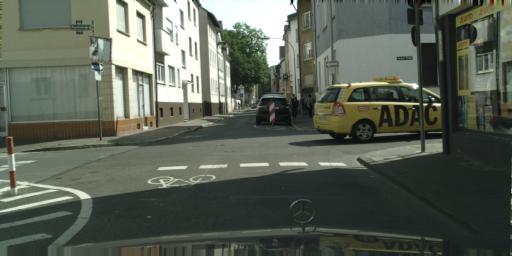}}}
	\subfigure[AFMSAT (Correl) ]{
		{\includegraphics[width=0.45\linewidth]{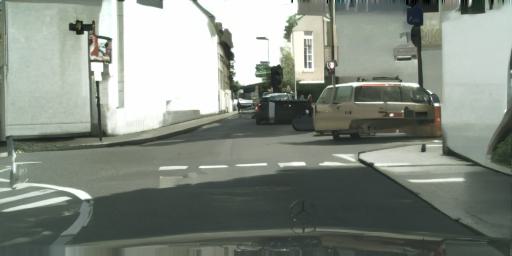}}}\\
  \subfigure[AFMSAT  ]{
		{\includegraphics[width=0.45\linewidth]{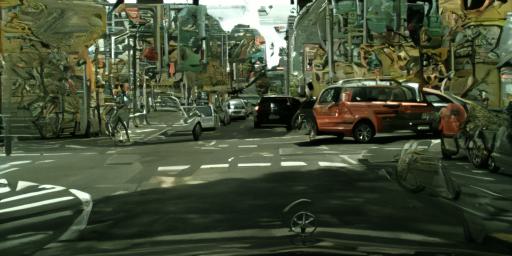}}}
  \subfigure[JSCC (Adapt)]{
		{\includegraphics[width=0.45\linewidth]{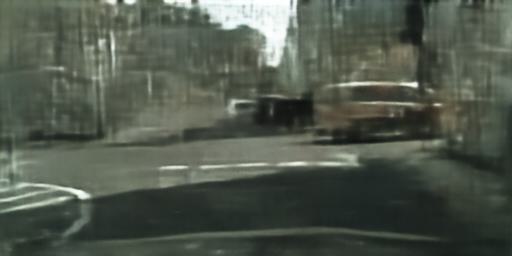}}}

	\caption{Image examples of the competing methods under bad channel conditions.}
	\label{AFMSAT_example}
\end{figure}

The conducted experiments affirm FMSAT's robust performance under varying SatCom channel conditions, characterized by rapid SNR fluctuations and CCI presence. Nonetheless, the efficacy of proposed methods is tested under extremely low SNRs and significant interference. Fortunately, satellites follow predictable trajectories, and noise along with interference from phenomena like rain attenuation and other devices can be somewhat estimated or forecasted. Thus, FM-based semantic approaches offer a promising strategy for prioritizing essential features under adverse conditions and leveraging commonalities from well-received images.


\subsection{Effectiveness of Adaption and Correlation}

Considering the impact of channel conditions on transmitter semantic features as known in advance, our proposed methods introduce a novel strategy that does not require additional bandwidth to mitigate CCI or to handle high noise levels. By computing the equivalent noises and affected semantic features under current satellite scenarios, we determine an adaptive transmission design for enhanced performance. When channel conditions are unfavorable, some semantic features may be omitted, necessitating an initial comparison of semantic segmentation performance.
 

  	\begin{table}[!t]
	\centering	
	\footnotesize
	\caption{The Ploss performance of the different methods using semantic segmentation. }

	\begin{tabular}{ccccc}
		\toprule
			Ploss& JSCC&FMSAT&FMSAT&FMSAT\\
   (SNR)& &(Difu)&(UNet+Difu)&(SegGPT+Difu)\\\midrule
-10 dB&0.559&0.423&0.405&0.396\\\midrule
-5 dB&0.388&0.338&0.340&0.341 \\ \midrule
0 dB&0.278& 0.267 &0.270&0.265\\ \midrule
5 dB&0.252&0.231&	0.253 &0.249\\ \midrule
0 dB&0.251& 0.220 &0.254&0.250\\
		
		\bottomrule
	\end{tabular}
	\label{Metric1}
\end{table}

The Ploss of the entire image does not solely reflect the superiority of adaptive methods. Furthermore, our FMSAT can opt to transmit either the entire image or only its crucial parts based on the channel conditions, a variation we term AFMSAT. We calculate the required Ploss between the important parts of an image, masking the other sections with zeros. Fig. \ref{ADA_Per} displays the required Ploss performance of various methods. AFMSAT(Correl) utilizes a previously well-received image for reconstruction, achieving optimal performance at low SNRs. As shown in Fig. \ref{ADA_Per}(a), AFMSAT consistently outperforms both FMSAT and JSCC methods by adapting its focus to crucial parts under low SNR and aiming to restore the entire image at high SNR. JSCC(Adapt) maintains moderate performance in required Ploss by always transmitting key parts, yet its performance drastically declines with decreasing SNR. This trend is more pronounced in the presence of CCI, as illustrated in Fig. \ref{ADA_Per}(b). Nonetheless, AFMSAT's performance slightly lags behind FMSAT when SNR exceeds 5 dB due to the CCI-specific encoder being optimized for 0 dB SNR, a limitation that could be addressed by training an additional encoder for higher SNRs.

Fig. \ref{AFMSAT_example} showcases examples at -5 dB SNR with 0.5 CCI. The figure reveals that, while JSCC(Adapt) can discern objects by sacrificing background detail, AFMSAT further enhances resolution and fills missing parts, rendering the received image sufficiently detailed for specific applications, such as remote driving. To supplement accurate background information (e.g., weather, architectural style), AFMSAT(Correl) combines efforts with a previous image, resulting in a restored image that surpasses other methods in quality.
 
These experiments underscore our method's efficacy in leveraging known channel conditions and well-received images, even in complex satellite scenarios. Despite this, retransmission remains inevitable. Our semantic approach can mend damaged segments and prioritize essential parts, rendering traditional error detectors suboptimal. Given that transmission via a regenerative satellite constitutes a multi-hop process, semantic errors should be identified at the satellite, where computational resources are scarce. The subsequent subsection will discuss the proposed error detector tailored for SatComs.

\subsection{Performances of Proposed Error Detector}

\begin{figure}[h]
	\centering
	\subfigure[]{
		{\includegraphics[width=0.98\linewidth]{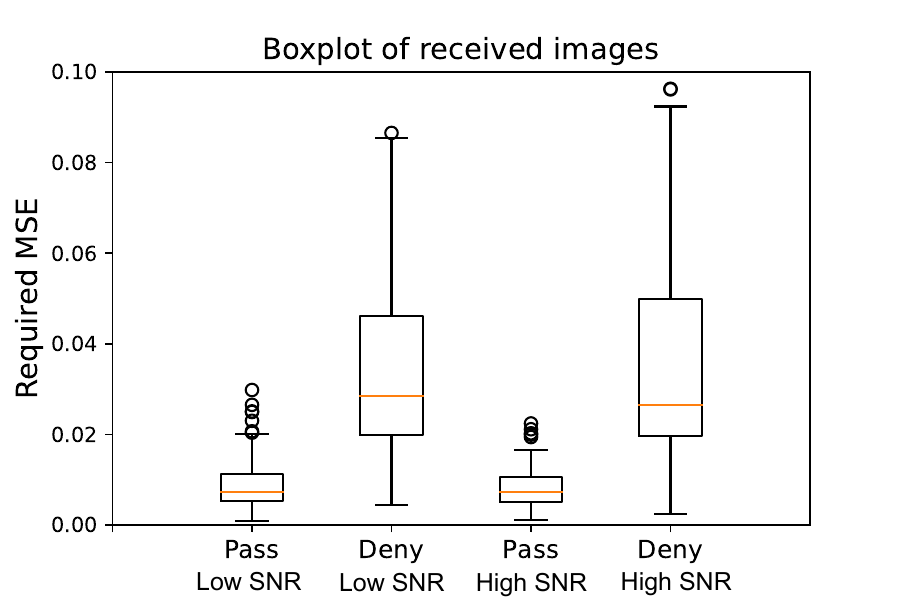}}}
\subfigure[]{
		{\includegraphics[width=0.98\linewidth]{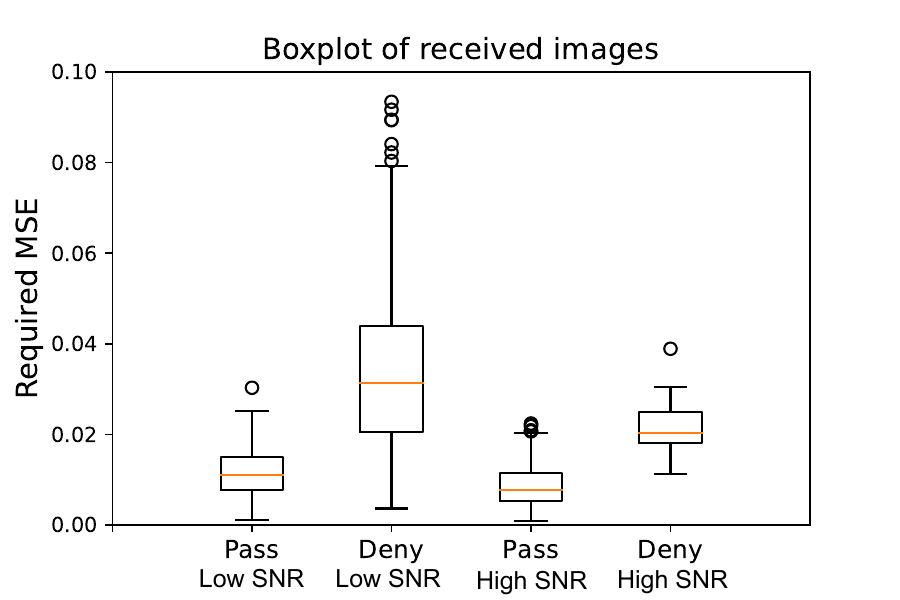}}}
	\caption{MSE performance of the required image part restored by the proposed methods. (a) The received images at the satellite. (b) The received images at the gateway.}
	\label{Err_det}
\end{figure}

This subsection delves into the performance of the proposed error detector within SatComs, exploring the detection process from the UT through to the gateway. Initially, images are transmitted from the UT, with codewords regenerated at the satellite before potential forwarding to the gateway, contingent on error detection outcomes. CCI conditions for both UT-satellite and satellite-gateway links simulate 0-0.5 affected subchannels. Low SNR scenarios imply transmission SNRs ranging from -10 to 0 dB, whereas high SNR scenarios span from 0 to 10 dB. The AFMSAT incorporates the proposed error detector, whereas the JPEG+LDPC model employs a CRC32 error detector.

\begin{figure}[h]
	\centering
	\subfigure[]{
		\label{Diff_Re_ACC}
		{\includegraphics[width=0.98\linewidth]{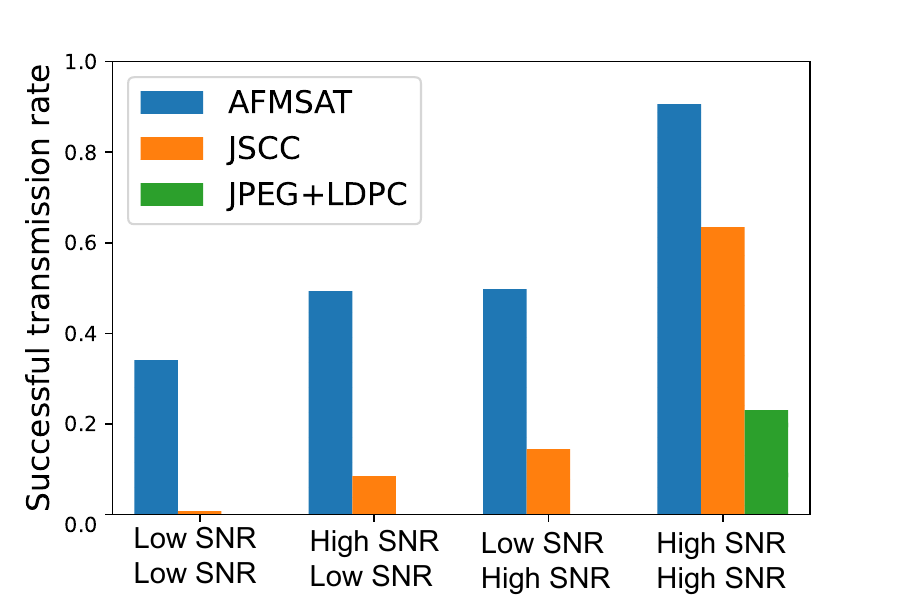}}}\\

\subfigure[]{
		\label{Diff_Re_MSE}
		{\includegraphics[width=0.98\linewidth]{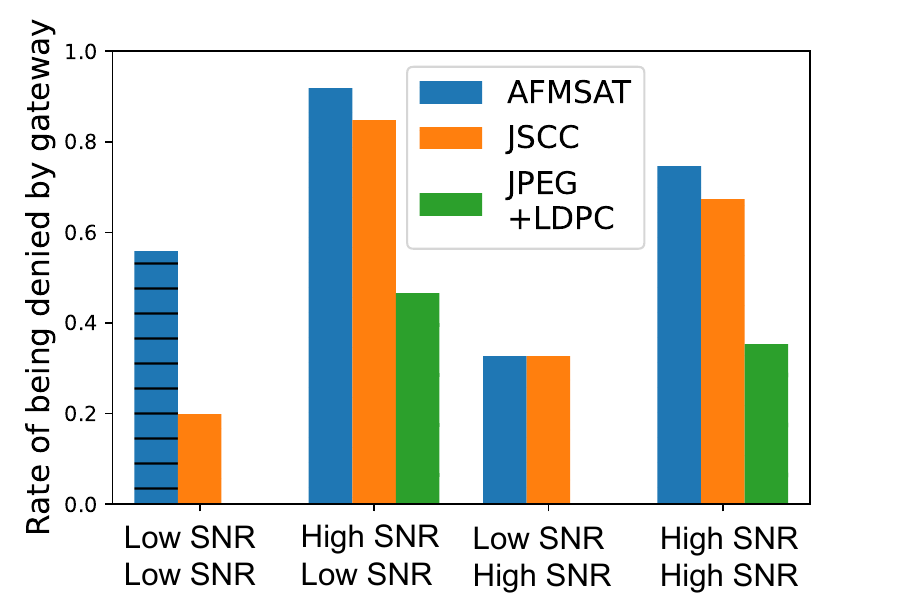}}}
	\caption{(a) Successful transmission rate at the gateway. (b) The proportion of the images detected by gateway to the total error images.}
	\label{Err_th}
\end{figure}

The box plots in Fig. \ref{Err_det} illustrate the MSE distribution for images acknowledged using the proposed method. Fig. \ref{Err_det}(a) focuses on the UT-satellite link, highlighting instances without detected errors under low SNRs, termed pass (lowSNR). The method leverages a 32-bit parity code to ensure acceptable MSE for reconstructed images from codewords, albeit with occasional misjudgments near the decision threshold. However, most outcomes are favorable, with acknowledged images typically featuring an MSE under 0.03. A similar pattern is observed in Fig. \ref{Err_det}(b) for images proceeding from the satellite to the gateway, with denied images under high SNRs presenting an MSE smaller than 0.04, indicating minimal transmission errors introduced in the satellite-gateway link.
 
Fig. \ref{Err_th}(a) assesses the success rate of transmitting 10,000 diverse images from the UT to the gateway, marking channel conditions on the x-axis. Notably, AFMSAT consistently ensures a high success rate across varying noise levels, outperforming the JSCC, especially as noise intensifies. The conventional JPEG+LDPC method depends on favorable SNRs, successfully transmitting images only when both links experience high SNRs.
 
Fig. \ref{Err_th}(b) evaluates the effectiveness of the satellite's rough detector by computing the ratio of images denied by the gateway to the total non-acknowledged images. Semantic methods, including AFMSAT and JSCC, refrain from correcting codewords at the satellite due to computational constraints. Consequently, transmission errors accumulating from the UT to the gateway often result in image degradation, particularly when the satellite-gateway link has low SNR. Conversely, the JPEG+LDPC method struggles with low SNR channels, with all images denied at the satellite under such conditions. Overall, the proposed error detection strategy proves efficient for satellite feedback, facilitating reduced transmission delays in HARQ systems.
 
\begin{table}[!t]
	\centering	
	\footnotesize
	\caption{Ploss performances of the images successfully accepted by the gateway.}

	\begin{tabular}{ccccc}
		\toprule
			UT-Satellite& \multicolumn{2}{c}{Low SNR}&\multicolumn{2}{c}{High SNR}\\ \midrule
   Satellite-Gateway& Low SNR&High SNR&Low SNR&High SNR\\\midrule
AFMSAT&0.203&0.185&0.181&0.119\\\midrule
FMSAT&0.130&0.133&0.178&0.143 \\ \midrule
JSCC&0.325&0.278&0.259&0.250\\ \midrule
JPEG+LDPC&/&/&	/ &0\\ \bottomrule
	\end{tabular}
	\label{Metric2}
\end{table}

An examination of the Ploss of images received at the gateway, as shown in TABLE \ref{Metric2}, reveals that the JPEG+LDPC method achieves a Ploss of 0 under high SNRs but fails to accept any images under adverse conditions, rendering the Ploss effectively infinite. The FMSAT methods outperform JSCC, as the diffusion model-based semantic reconstruction adeptly mends damaged sections, provided the codeword is not severely compromised. While AFMSAT slightly underperforms FMSAT under challenging channel conditions due to the exclusion of non-essential parts, it compensates by enabling the transmission of a greater number of images compared to FMSAT.

\section{Conclusion}
\label{s6}

This paper presents a novel FM-based semantic framework for SatComs named FMSAT. This innovative approach effectively navigates the challenges posed by the high mobility and aggressive frequency reuse associated with LEO satellites. The core of the framework lies in its use of FM-based semantic segmentation and reconstruction networks, which safeguard critical features and bolster visual quality for image transmission from the UT to the gateway. 
A significant advantage of this method is its ability to leverage known channel conditions and previously well-received images to refine performance, thus alleviating the need for satellites to frequently adjust bandwidth usage. Semantic methods provide the flexibility to disregard certain non-essential parts of an image under adverse channel conditions. Moreover, semantic reconstruction, when combined with a well-received image, adeptly compensates for missing segments, ensuring the integrity of the transmitted image.

Crucially, the transmission of semantic features with errors necessitates a novel error detection mechanism tailored for regenerative satellites. The proposed error detector, requiring only a brief parity code alongside the received codeword at the satellite, efficiently identifies most errors in situ. This capability is instrumental in minimizing transmission delays, ensuring that corrective actions can be initiated promptly and effectively.
Overall, the FMSAT framework introduces a significant advancement in SatComs, offering a robust solution to the complexities posed by LEO satellite operations. 



	\bibliographystyle{IEEEtran}
	\bibliography{bibtex0320}
	
	%
	
	
	
\end{document}